\magnification1200
{\settabs 2\columns
\+Preprint n. &Dipartimento di Fisica\cr
\+&Universit\`a di Roma ``La Sapienza''\cr
\+&I.N.F.N. - Sezione di Roma\cr}
\vskip 2cm
\centerline{{\bf FUNCTIONAL ORDER PARAMETERS}
\footnote{}{Dedicated to Hiroomi Umezawa}}
\centerline{{\bf FOR THE QUENCHED FREE ENERGY}
\footnote{}{}}
\centerline{{\bf IN MEAN FIELD SPIN GLASS MODELS}\footnote{$\ddag$}{Research
supported in part by MURST (Italian Minister of University and Scientific and
Technological Research) and INFN (Italian National Institute for Nuclear
Physics).}}  \smallskip 
\centerline{by}
\smallskip \centerline
{Francesco Guerra}
\centerline{Dipartimento di Fisica, Universit\`a di Roma ``La Sapienza'',}
\centerline{and Istituto Nazionale di Fisica Nucleare, Sezione di Roma,}
\centerline{Piazzale Aldo Moro, 2, I-00185 Roma, Italy.}
\centerline{e-mail guerra@roma1.infn.it}
\bigskip\bigskip
\centerline{December 1992}
\vfill
\beginsection ABSTRACT.

In the Sherrington-Kirkpatrick mean field model
for spin glasses, we show that the quen\-ched average of the free energy 
can be expressed through a couple of functional order parameters, in a form
very similar to the one found in the frame of the
replica symmetry breaking method. The functional order parameters are 
implicitely given in terms of fluctuations of thermodynamic variables.

Under the assumption that the two order parameters can be chosen to be the
same, in the thermodynamic limit, it is shown that the Parisi free energy
is a rigorous upper bound for the free energy of the model.    
\vfill\eject

Let us introduce the partition function $Z_N(\beta,J)$ and the
free energy $F_N(\beta,J)$ for the Sherrington-Kirkpatrick mean field spin glass model [1,2]
in the form
$$Z_N(\beta,J)=
\sum_{\sigma_1\dots\sigma_N}\exp({\beta\over\sqrt{N-1}}\sum_{(i,j)}J_{ij}\sigma_i\sigma_j)=
\exp(-\beta F_N(\beta,J)).
\eqno(1)$$
The $\sigma$'s are Ising spins describing a generic configuration of the system 
$$\sigma:\quad\{1,2,\dots,N\}\ni i\to\sigma_i\in Z_2=\{-1,1\}.
\eqno(2)$$
For each of the $N(N-1)/2$ couples of sites
$(i,j)$, $i\ne j$, over which the sum $\sum_{(i,j)}$ runs, we have introduced independent random
variables $J_{ij}=J_{ji}$, $i\ne j$,  identically distributed , called quenched variables. The
$\sigma$'s are mesoscopic random variables subject to thermodynamic
equilibrium. The $J$'s do not participate to thermodynamic equilibrium,
but act as a kind of random environment on the $\sigma$'s. For the sake
of simplicity, we assume that the $J$'s have unit Gaussian distribution
with
$$E(J_{ij})=0,\quad E(J_{ij}^2)=1,
\eqno(3)$$
where $E$ denotes averages with respect to the $J$ variables. The parameter 
$\beta$ is the inverse temperature in proper units.

We are interested in the expression of the thermodynamic limit $N\to\infty$ for the free energy per
spin, averaged over the external noise (quenched average),
$$\lim_{N\to\infty}N^{-1}E\bigl(\log
Z_N(\beta,J)\bigr). 
\eqno(4)$$

Let us introduce the marginal free energy, {\sl i.e.} the increment in the free energy when an
additional $(N+1)$th spin is added to a system of $N$ spins, at the {\sl same}
inverse temperature 
$$-\beta\bigl(F_{N+1}(\beta,J)-F_N(\beta,J)\bigr)=
\log Z_{N+1}(\beta,J)-\log Z_N(\beta,J).
\eqno(5)$$
Then we have
\smallskip\noindent
{\bf Proposition 1.} The quenched average of the marginal free energy and the free energy per
spin can be expressed in the following form
$$E\bigl(\log Z_{N+1}(\beta,J)\bigr)-E\bigl(\log Z_N(\beta,J)\bigr)=
log 2 + \psi_N(\beta) - \phi_N(\beta),     
\eqno(6)$$
$$(N+1)^{-1}E\bigl(\log Z_{N+1}(\beta,J)\bigr)=
log 2 + {\overline\psi}_N(\beta) - {\overline\phi}_N(\beta),
\eqno(7)$$
$$\psi_N(\beta)=
E\log\omega_{N}(\cosh{\beta\over\sqrt N}\sum_i J_i\sigma_i),
\eqno(8)$$
$$\phi_N(\beta)=
E\log\omega_{N}(\exp{\beta\over\sqrt{N(N-1)}}\sum_{(ij)}{\tilde J}_{ij}\sigma_i\sigma_j),
\eqno(9)$$
$${\overline\psi}_N(\beta)=
(N+1)^{-1}\sum_{K=0}^{N}\psi_K(\beta),\quad\quad
{\overline\phi}_N(\beta)=
(N+1)^{-1}\sum_{K=0}^{N}\phi_K(\beta).
\eqno(10)$$

Here $\omega_{N}$ is the Boltzmann state with {\sl Boltzmannfaktor} as in (1) with $\sqrt {N-1}$
replaced by $\sqrt N$, and the $J_i$'s and ${\tilde J}_{ij}$'s, $i,j=1,2,\dots,N$, $i\ne j$, are
$N$ and $N(N-1)/2$, respectively, independent random variables, with unit Gaussian distribution.
We call $J_{ij}$ the stale noise, and $J_i$ and ${\tilde J}_{ij}$ the fresh
noise.

For the proof we can write
$$E\bigl(\log Z_{N+1}(\beta,J)\bigr)=
E\log\sum_{\sigma_1\dots\sigma_{N+1}}\exp({\beta\over\sqrt{N}}
\sum_{(i,j)}^{1\dots N}J_{ij}\sigma_i\sigma_j+\sum_i^{1\dots N} J_{iN+1}\sigma_i\sigma_{N+1})=$$
$$=\log2 + E\log\omega_{N}(\cosh{\beta\over\sqrt N}\sum_i J_i\sigma_i)+
E\log\sum_{\sigma_1\dots\sigma_N}\exp({\beta\over\sqrt{N}}
\sum_{(i,j)}J_{ij}\sigma_i\sigma_j),
\eqno(11)$$
where we have explicitely performed the sum over $\sigma_{N+1}$ and  have called $J_i$ the old
$J_{iN+1}$.

Let us now consider
$$E\log Z_N(\beta,J)=E\log
\sum_{\sigma_1\dots\sigma_N}\exp({\beta\over\sqrt{N-1}}\sum_{(i,j)}J_{ij}\sigma_i\sigma_j).
\eqno(12)$$
By introducing a fresh set of independent noise ${\tilde J}_{ij}$, with the same normalization
as in (3), we can replace $J_{ij}/\sqrt{N-1}$ with the stochastically equivalent sum
$J_{ij}/\sqrt N + {\tilde J}_{ij}/\sqrt{N(N-1)}$, in fact the two random variables have the
same mean and tha same covariance. Therefore, we have
$$E\log Z_N(\beta,J)=E\log
\sum_{\sigma_1\dots\sigma_N}\exp({\beta\over\sqrt{N}}\sum_{(i,j)}J_{ij}\sigma_i\sigma_j+
{\beta\over\sqrt{N(N-1)}}\sum_{(ij)}{\tilde J}_{ij}\sigma_i\sigma_j)=$$
$$=E\log\omega_{N}\bigl(\exp({\beta\over\sqrt{N(N-1)}}
\sum_{(ij)}{\tilde J}_{ij}\sigma_i\sigma_j)\bigr)+
E\log\sum_{\sigma_1\dots\sigma_N}\exp({\beta\over\sqrt{N}}
\sum_{(i,j)}J_{ij}\sigma_i\sigma_j),
\eqno(13)$$
and (6) follows. Now we can write (6) for a generic $K$, and sum from $K=0$ to $K=N$. With the
obvious notations
$$E\log Z_0(\beta,J)=0,\quad\quad \psi_0(\beta)=0,\quad\quad \phi_0(\beta)=0,  
\eqno(14)$$
we immediately have (7).

Useful information on the functions $\psi_N(\beta)$, ${\overline\psi}_N(\beta)$,
$\phi_N(\beta)$, ${\overline\phi}_N(\beta)$ is given by 
\smallskip\noindent
{\bf Theorem 2.} The following bounds hold
$$\int\log\cosh(\beta z)\ d\mu(z)\le\psi_N(\beta),\ {\overline\psi}_N(\beta)
\le {\beta^2}/2,
\eqno(15)$$
$$0\le \phi_N(\beta),\ {\overline\phi}_N(\beta)
\le {\beta^2}/4,
\eqno(16)$$
where $d\mu(z)=\exp(-z^2/2)\ dz/\sqrt{2\pi}$ is the unit Gaussian distribution.

For $\psi_N(\beta)$ the proof has been given in [3]. It is based on either annealing the $E$
averages (upper bound), or quenching the $\omega$ averages (lower bound). The bound for
${\overline\phi}_N(\beta)$ follows easily from the definition (10). In the same way one proves
(16).

Let us now introduce the convex set  $\cal X$ of functional order parameters of
the type 
$$x:\quad[0,1]\ni q\to x(q)\in[0,1],
\eqno(17)$$
with the $L^1(dq)$ distance norm. We induce on $\cal X$ a partial ordering, by defining
$x\le{\bar x}$ if 
$x(q)\le{\bar x}(q)$, for all $0\le q\le1$, and introduce the extremal order parameters
$x_0(q)\equiv 0$ and
$x_1(q)\equiv1$, such that for any $x$ we have $x_0(q)\le x(q)\le x_1(q)$.

For each $x$ in $\cal X$, and $\beta\ge0$, let us define the function with values
$f(q,y;x,\beta)$, $0\le q\le1$, $y\in R$, 
as the solution of the nonlinear
antiparabolic equation 
$$\partial_q f+{1\over2}\bigl(f^{\prime\prime}+x(q){f^\prime}^2\bigr)=0,
\eqno(18)$$
with final condition
$$f(1,y;x,\beta)=\log\cosh(\beta y).
\eqno(19)$$
In (18), $f^\prime=\partial_y f$ and $f^{\prime\prime}=\partial_y^2 f$.

As a shorthand notation, for each $x$ in $\cal X$, and $\beta\ge0$, we define at $q=0$,
$y=0$
$$f(x,\beta)=f(0,0;x,\beta).
\eqno(20)$$

In Ref. [3], we have shown that (18,19) arise in a very natural way as a result of exact
corrections to the annealing approximation $\log E\omega(\dots)$ in the evaluation of the quenched
average in (8). 

The following theorem summarizes some important properties [3] of $f(x,\beta)$.
\smallskip\noindent
{\bf Theorem 3.} The function $f(x,\beta)$ is monotone in $x$, {\sl i.e.} $x\ge \bar x$ implies
$f(x,\beta)\ge f(\bar x,\beta)$. Moreover, the following bounds hold
$$f(x_0,\beta)=\int\log\cosh(\beta z)\ d\mu(z)\le f(x,\beta)
\le {\beta^2}/2=f(x_1,\beta).
\eqno(21)$$

In [3], we have also proven the following representation theorem.
\smallskip\noindent
{\bf Theorem 4.} There exists a nonempty hypersurface $\Sigma_N(\beta)$ in $\cal X$, such that, for
any $x\in \cal X$ and $f$ solution of (18,19), we have the following representation
$$\psi_N(\beta)=f(x,\beta).
\eqno(22)$$
Any  family of functional order parameters, $x_{\epsilon}$,
depending continuously in the $L^1$ norm on the variable $\epsilon$,
$0\le\epsilon\le1$, with $x_0\equiv0$, and $x_1\equiv1$, and nondecreasing in
$\epsilon$ must necessarily cross $\Sigma_N(\beta)$ for some value of the variable $\epsilon$ (we
say that $\Sigma_N(\beta)$ has the monotone intersection property).

By using the same method, we can easily prove the following easy generalization.
\smallskip\noindent
{\bf Theorem 5.} The average ${\overline\psi}_N(\beta)$, defined in (10), admits also a
representation
$${\overline\psi}_N(\beta)=f(x,\beta),
\eqno(23)$$
for $x$ on some hypersurface $\Sigma_N(\beta)$ (which will be in general sligthly different from
the hypersurface appearing in the previous (22)).

The method of Ref. [3] allows to give implicit expressions for the elements of $\Sigma_N(\beta)$ in
terms of fluctuations, but the very existence of $\Sigma_N(\beta)$, with 
the monotone intersection property, follows from a very simple argument. In fact, from the bounds
(15) and (21), and the monotonicity of $f(x,\beta)$ in $x$, given by Theorem 3, we immediately
have the existence of a nonempty $\Sigma_N(\beta)$.

Similar representation formulae hold for $\phi_N(\beta)$ and  ${\overline\phi}_N(\beta)$.
\smallskip\noindent
{\bf Theorem 6.} There exist  nonempty convex linear sets $\tilde{\Sigma}_N^{\prime}(\beta)$ and 
$\tilde{\Sigma}_N(\beta)$ in $\cal X$, such that
$$\phi_N(\beta)\ or\ {\overline\phi}_N(\beta)=
{1\over2}\beta^2\int_0^1 q\ \tilde x(q)\ dq,
\eqno(24)$$
for any $\tilde x\in \tilde{\Sigma}_N^{\prime}(\beta)$, or 
$\tilde x\in \tilde{\Sigma}_N(\beta)$, respectively.

The proof follows from a simple cumulant expression. Let us introduce the interpolating
parameter $q$, $0\le q \le1$, and define
$$\phi(q)=
E\log\omega_{N}(\exp{{\beta q}\over\sqrt{N(N-1)}}\sum_{(ij)}{\tilde J}_{ij}\sigma_i\sigma_j),
\eqno(25)$$
so that $\phi(0)=0$ and $\phi(1)=\phi_N(\beta)$, as defined in (9). Let us take the derivative
$${d\over {dq}}\phi(q)={\beta\over\sqrt{N(N-1)}}\sum_{(ij)}E\bigl({\tilde J}_{ij}
\omega_N^{-1}(\exp(\dots))
\omega_N(\sigma_i\sigma_j \exp(\dots))\bigr). 
\eqno(26)$$
Then we can exploit the general integration by parts formula
$$E\bigl({\tilde J}_{ij}F(J)\bigr)=E\bigl({\partial\over{\partial{\tilde J}_{ij}}}F(J)\bigr),
\eqno(27)$$
and obtain
$${d\over{dq}}\phi(q)={1\over2}\beta^2 {\tilde x}(q),       
\eqno(28)$$
where
$${\tilde x}(q)=1-{2\over\sqrt{N(N-1)}}\sum_{(ij)}
E\Bigl({{\omega_{N}^2(\sigma_i\sigma_j\exp{{\beta
q}\over\sqrt{N(N-1)}}\sum_{(ij)}{\tilde J}_{ij}\sigma_i\sigma_j)}\over
{\omega_{N}^2(\exp{{\beta
q}\over\sqrt{N(N-1)}}\sum_{(ij)}{\tilde J}_{ij}\sigma_i\sigma_j)}}\Bigr).
\eqno(29)$$
Clearly, we have the inequality
$$0\le {\tilde x}(q)\le1.
\eqno(30)$$
By integrating (28) on $dq$ we have the representation (24) for $\phi_N(\beta)$. This shows that
$\tilde{\Sigma}_N^{\prime}(\beta)$ is nonempty, because $\tilde x$ is explicitely defined by
(29). Of course, all functional order parameters, which give the same value for the integral in
(24), are acceptable. This is how the convex linear set $\tilde{\Sigma}_N^{\prime}(\beta)$
arises. Also in this case we have the monotone intersection property. The representation (24)
for ${\overline\phi}_N(\beta)$ follows easily from the definition (10).

Let us also explicitely remark that the representations given in Theorems 4,5,6 hold for any
even state $\omega_N$, not necessarily as that arising in (8,9). Of course, the involved
hypersurfaces do depend on the particular $\omega_N$.

By collecting all results of Theorems 5 and 6, and the definition (7), we have the following
basic representation theorem for the quenched average of the free energy per spin
\smallskip\noindent
{\bf Theorem 7.} There exist  nonempty hypersurfaces $\Sigma_N(\beta)$ 
and $\tilde{\Sigma}_N(\beta)$ in $\cal X$, such that
$$(N+1)^{-1}E\bigl(\log Z_{N+1}(\beta,J)\bigr)=
\log2+f(x,\beta)-{1\over2}\beta^2\int_0^1 q\ {\tilde x}(q)\ dq,
\eqno(31)$$
for any $x\in \Sigma_N(\beta)$ and ${\tilde x}\in \tilde{\Sigma}_N(\beta)$.
Elements of these two hypersurfaces can be expressed implicitely in terms of fluctuations.

The representation (31) is equivalent and complementary to the representation given in [2],
which  involves the order parameter $x$ for {\sl different} values of
$\beta$. Here {\sl two} order parameters are involved, but at the {\sl same} value of
$\beta$.

This representation is very similar to that found  in the frame of the
replica symmetry breaking method, with Parisi {\sl Ansatz} [2], where the two order
parameters are considered to be the same, at least in the thermodynamic limit. Therefore,
we are led to explore the consequences of the following
\smallskip\noindent
{\bf Assumption 8.} Let $a_N(\beta)$ be the $L^1$ distance between the
hypersurfaces $\Sigma_N(\beta)$ 
and ${\tilde x}\in \tilde{\Sigma}_N(\beta)$
$$a_N(\beta)=\inf \int^1_0|x(q)-{\tilde x}(q)|\ dq,\ x\in \Sigma_N(\beta),\ 
{\tilde x}\in \tilde{\Sigma}_N(\beta),
\eqno(32)$$
and assume
$$\lim_{N\to \infty}a_N(\beta)=0.
\eqno(33)$$

Let us also define the Parisi free energy $f_P(\beta)$ at inverse temperature $\beta$ as
$$-\beta f_P(\beta)= \inf_{x\in {\cal X}} \bigl(\log2+f(x,\beta)-
{1\over2}\beta^2\int_0^1 q\  x(q)\ dq \bigr).
\eqno(34)$$
Then we have
\smallskip\noindent
{\bf Proposition 9.} Under the stated assumption, in the thermodynamic limit, we have
$$\liminf_{N\to \infty} (N+1)^{-1}E\bigl(\log Z_{N+1}(\beta,J)\bigr) \ge
-\beta f_P(\beta).                
\eqno(35)$$
The proof is immediate. In fact, for the r.h.s. of (31) we have
$$\log2+f(x,\beta)-{1\over2}\beta^2\int_0^1 q\  x(q)\ dq+
{1\over2}\beta^2\int_0^1 q\  (x(q)-{\tilde x}(q))\ dq \ge
-\beta f_P(\beta)-a_N(\beta),
\eqno(36)$$
and the result follows by taking the limit $N\to\infty$.

Therefore, the Parisi free energy, with these assumptions, is proven to be at least a rigorous
upper bound for the infinite volume limit of the free energy of the model.

In a forthcoming paper [4], we show that there is good evidence, not a definite mathematical
proof as yet, that the two order parameters in (31) can be taken the same, in the thermodynamic
limit, and moreover that the Parisi free energy is the true free energy, and not only an upper
bound.
\bigskip\bigskip
\beginsection REFERENCES

\item{ [1]} D. Sherrington and S. Kirkpatrick: Solvable model of a spin glass,
Phys. Rev. Lett., {\bf35}, 1792 (1975).
\item{ [2]} M. M\'ezard, G. Parisi, and M. A. Virasoro: {\sl Spin Glass Theory
and Beyond}, World Scientific, Singapore, 1987, and reprints included there.
\item{[3]} F. Guerra: Fluctuations and Thermodynamic Variables in Mean Field Spin Glass Models,
in: {\sl Stochastic Processes, Physics and Geometry}, S. Albeverio {\sl et al.},
eds, World Scientific, Singapore, 1992.
\item{[4]} F. Guerra: On the mean field spin glass model, in
preparation. \vfill\eject\bye